# Identifying potential 'breakthrough' publications using refined citation analyses: Three related explorative approaches

Jesper W. Schneider*
Danish Centre for Studies in Research & Research Policy,
Department of Political Science & Government
Aarhus University
Bartholins Allé 7, DK-8000 Aarhus C
Denmark
jws@ps.au.dk

Rodrigo Costas
Centre for Science and Technology Studies, CWTS
Leiden University
Wassenaarseweg 62A, 2333 AL Leiden,
The Netherlands
rcostas@cwts.leidenuniv.nl

*corresponding author

## Abstract

The article presents three advanced citation-based methods used to detect potential breakthrough papers among very highly cited papers. We approach the detection of such papers from three different perspectives in order to provide different typologies of breakthrough papers. In all three cases we use the classification of scientific publications developed at CWTS based on direct citation relationships. This classification establishes clusters of papers at three levels of aggregation. Papers are clustered based on their similar citation orientations and it is assumed that they are focused on similar research interests. We use the clustering as the context for detecting potential breakthrough papers. We utilize the Characteristics Scores and Scales (CSS) approach to partition citation distributions and implement a specific filtering algorithm to sort out potential highly-cited 'followers', papers not considered breakthroughs in themselves. After invoking thresholds and filtering, three methods are explored: A very exclusive one where only the highest cited paper in a micro-cluster is considered as a potential breakthrough paper (M1); as well as two conceptually different methods, one that detects potential breakthrough papers among the two percent highest cited papers according to CSS (M2a), and finally a more restrictive version where, in addition to the CSS two percent filter, knowledge diffusion is also taken in as an extra parameter (M2b). The advance citation-based methods are explored and evaluated using specifically validated publication sets linked to different Danish funding instruments including centres of excellence.

## Introduction

The conception of 'breakthrough' research is typically linked to discovery and often characterized as creative, transformative and ground breaking research (Häyrynen, 2007). Breakthrough research is for example coupled with solutions to broad and complex research problems, or challenging established theories and scientific paradigms, or establishing fundamental new ways of using methods and instruments, or sometimes interdisciplinary integration of different research perspectives. In that respect, setting out to do breakthrough research is often also seen as having a high risk of failure. From a science policy perspective, attention has been given at the national and international levels to setting up funding instruments and research infrastructures with the intention to motivate and foster breakthrough research (Häyrynen, 2007; Öquist & Benner, 2012). An important challenge here is the initial selection and funding of proposals that may or may not end up as being characterized as breakthrough research. From a science study perspective, questions concerning what psychological, organizational and institutional factors influence, and what funding schemes would best support, breakthrough research, have been and continue to be of interest (e.g., Simonton, 1988; Hollingsworth, 2002; Simonton, 2004; Heinze et al., 2007; Heinze, 2008; Heinze et al., 2009; Öquist & Benner, 2012; Winnink, Tijssen & van Raan, 2015); as well as more predictive approaches where early identification of potential breakthrough research are modelled (e.g., Simonton, 1988, 2004; Chen et al., 2009; Small & Klavans, 2011; Chen, 2012; Ponomarev et al., 2014a; Ponomarev et al., 2014b).

To early sociologists of science, breakthroughs were conceptualized in relation to the reward system in science (Merton, 1957). Emphasis is on the novelty of research results and the important concept here is 'originality' which the scientist's reputation rests on and on the basis of which they are rewarded. Originality is seen as a flexible concept which can refer, for example, to the creation of new theories, to the improvement of an existing theory or to providing a better description of a known phenomenon. But it is also a concept which is constrained by the norms of science as argued by Merton (1973). Kuhn's perspective is different; here breakthroughs are seen as fundamental paradigmatic breaks with norms and linked to scientific revolutions (Kuhn, 1996). Kuhn's focus is the natural sciences where breakthroughs in his conception are rare events. But Kuhn also maintains that in normal science, breakthroughs and revolutions leading to paradigm shifts have important and mutually complementary roles within the system of science. Galison (1997) disputes Kuhn's ideas about incommensurability and how science lurches from one paradigm to the next. Using the 'trading zone' metaphor, Galison argues that different material cultures can meet and exchange 'goods' in these zones without sharing beliefs or values. In his case study of particle physics he shows that when 'goods' were exchanged shifts of meaning took place, thus allowing two communities, experimentalists and theoreticians, to collaborate corresponding to small breakthroughs and not paradigm shifts. Inspired by Kuhn, later sociologists of science see radical breakthroughs as challenging the power structures within the institutions of science, and as such 'novelty' in this sense is feared and to some extent countered by the 'ruling classes' in a domain, for example through the peer review process (Bourdieu, 1988; Becher, 1989; Shatz, 2004). Contrariwise, quantitative models by De Solla Price (1975), Bruckner, Ebeling and Scharnhorst (1990) and later van Raan (2000), portray science as a complex, largely self-organizing 'cognitive eco-system' encompassing many self-similar subsystems or research fields where each field in principle originates from an important breakthrough in scientific research and each, in turn, gives rise to other breakthroughs, new research fields, with a probability proportional to its size. This is clearly in opposition to Kuhn's model of scientific development.  In this conception, and in line with Galison (1997), science does not evolve by dramatic, socially and psychologically driven paradigm shifts marking the transition from revolutionary to normal cumulative scientific activity (De Bellis, 2009).  In van Raan's view, new ideas and research fields, which may well be deemed revolutionary in their own right, although on different scales of magnitude, develop almost linearly from antecedent ones, so that "there is no 'normal' science alternated with well-defined periods of 'revolutionary' science in which new paradigms start to dictate the rules. Science is always revolutionary, but by the typical statistics of complex systems, there are mostly smaller and only rarely big breakthroughs" (van Raan, 2000, p. 360).

It is evident that these different conceptions of breakthrough research have different aims and address the issue form different levels, from Kuhn's all-encompassing rare paradigm-shifts to Merton's and van Raan's more day-to-day and smaller scale incremental workings within scientific fields.  All seem to agree that 'original ideas' are what drives science and novelty seems to work and be acknowledged at different levels.  Endorsements, citations, rewards and prizes may be bestowed upon such research, but 'original ideas' and the potential breakthroughs produced by those ideas need to be sanctioned in order to be considered as such, and that process is often influenced more by scientific elites, social networks, political considerations, and the Matthew effect, than the idealized norms of science.

Using bibliometric data to identify or predict potentially 'excellent' or breakthrough research has been an aim for decades.  For example, Garfield and Welljamsdorof (1992) analysed whether rankings of highly cited authors confirmed or even predicted Noble prize awards.  Indeed, highly cited units of analysis (i.e., articles, authors, groups or institutions) have become a common denominator for excellence (e.g., Tijssen, Visser & van Leeuwen, 2002; Aksnes, 2003), and the target for predicting potential breakthrough research (e.g., Ponomarev et al., 2014b).  Identification and model building are typically retrospective in as much as excellent or breakthrough research is determined by other means than citation analyses and then from the citation patterns of these exemplars, comparisons and predictions are made.  In general, very highly cited units seem to be good predictors for prizes, awards and peer acknowledgement of excellence, although one can certainly not rule out a kind of circularity in such reasoning.

The aim of the present article is also to identify breakthrough research by focusing on highly cited articles. We fully recognize the dual nature of academic research as both a tool for incremental problem-solving within the basic knowledge framework and a way of seeking breakthroughs that generate new, unforeseen opportunities. Using highly cited publications as markers for potential breakthrough research unavoidably entails that we operate on different levels of potential breakthroughs and we basically cannot quantitatively differentiate between them. What follows is our pragmatic definition of potential breakthrough research in the present study.

To precisely define what could constitute a breakthrough paper or breakthrough research is very difficult and eventually turns out to be rather vague. For example, Karlsson and Persson (2012) simply define breakthrough papers as the 10 percent most cited publications in their respective fields. If we look for the most common dictionary definition of breakthrough (focusing on those meanings related to this study) we find the following definitions:

- Oxford dictionary[1]: "a sudden, dramatic, and important discovery or development".
- Collins Dictionary[2]: "a significant development or discovery".
- Dictionary.com[3]: "any significant or sudden advance, development, achievement, or increase, as in scientific knowledge or diplomacy that removes a barrier to progress".
- The Free dictionary[4]: "a major achievement or success that permits further progresses".

In general, we can see that there is no straightforward definition of what could constitute breakthrough research and the concepts used to define breakthrough are rather vague and difficult to operationalize. What eventually is considered breakthrough research is a matter to be decided by peers. However, based on all the previous discussions we can select a few characteristics that will help us frame and implement our methodology. These characteristics are the following:

- 'Major achievements' or the idea of 'significant and important advances that permit further progress' can be operationalized as articles that are (extremely) highly cited (i.e. they can be considered as major achievements from a bibliometric point of view) and have also had knowledge dissemination and influence within its own field but also in other fields

---

[1] http://oxforddictionaries.com

[2] http://www.collinsdictionary.com

[3] http://dictionary.reference.com/

[4] http://www.thefreedictionary.com/breakthrough

- (interdisciplinary spread), thus contributing to further progress within and outside the same field(s).
- 'Sudden or dramatic advance' can be operationalized as articles that are not simple 'followers' of other breakthroughs, but they are important publications on their own and therefore have a distinctive nature compared with previous articles.

Based on these rationales, we propose the following working definition of a breakthrough paper in this study:

"*a very highly cited paper, with an important citation spread over its own field and also other fields of science, and it must be a paper that is not a mere follower of other highly cited publication(s), it must have a genuine relevance on its own*".

Other definitions are of course conceivable and do not necessarily have to rely on citations, e.g. network statistics are an obvious alternative. We assume that identified potential breakthrough papers are proxies or markers of breakthrough research. A paper can indeed report what eventually turns out to be breakthrough research, but what turns out to be breakthrough research can also be the sum of knowledge claims in a number of papers, where some of them are perhaps not highly cited. It is therefore foolhardy to believe that a quantitative attempt at detecting breakthrough research with *one* specific approach can be exhaustive or flawless, this is clearly not the case, and we fully acknowledge the limitations of our citation based approach. But, as one modest attempt, citation analysis is an interesting approach to explore in this setting. If we assume that in the fields we analyse, research results are mainly reported in international journal articles, and if we also assume that within narrower research areas, highly cited publications to a large extent signal impact and use of the content in these papers by the research community, though noise will also be in there, then it would be reasonable to assume that potential breakthrough research in many instances would be reported in papers that subsequently become highly cited exactly because the research has breakthrough potential.

    These are the basic assumptions of this analysis and if they are accepted, three major methodological challenges remain: 1) establishing an exhaustive network of research areas in which breakthrough papers can be detected (establishing the context); 2) detecting potential breakthrough papers among the set of (extremely) highly cited papers (construction of advanced citation analyses); and 3) determine whether potential breakthrough papers are indeed proxies for some breakthrough research (evaluation of the citation-based methods).

The following section presents and discusses the data and methods behind the construction of the network of research areas; outlines the three advanced citation-based methods; and presents the data sets used for evaluation of the proposed methods. The following section presents the results, and finally we evaluate and discuss the results and reflect upon the utility of the proposed approach.

## Data and methods

The total data universe used for the present study is publications from 1993 to 2011 in the Web of Science (WoS) database of CWTS, Leiden University, Netherlands. The total WoS dataset is used to establish the networks and clusters of journal articles and to calculate global citation statistics. Within this confined data set, we have established four distinct sub-sets of validated journal articles which will be used for the specific analyses and comparisons. The analyses are focused on a set of WoS journal articles from 1993 to 2011 linked to 66 Danish Centres of Excellence (CoE) funded by the Danish National Research Foundation (DNRF). The CoE are the main unit of analysis and their publications are analysed in order to examine to what extent they qualify as potential breakthroughs given our operationalization. In a subsequent analysis, we merge CoE publications from 2005 to 2011 into a combined set linked to the aggregate DNRF funding instrument. Potential breakthrough papers from this aggregate DNRF publication set are compared to a corresponding set of validated publications linked to the Danish Council for Independent Research (DFF). The latter set is also aggregated as publications come from various different grant types funded by the DFF. Finally, we are also able to compare the two publication sets linked to the different Danish funding instruments to the total set of Danish publications in the same period.

*Establishing a network and clusters of papers*

We approach the detection of breakthrough papers from three different perspectives in order to provide different typologies of breakthrough papers. In all three cases we use the classification of scientific publications developed at CWTS (Waltman & van Eck, 2012). Analysing potential breakthrough research should commence in the local context of the research area or field where such knowledge claims are first proposed. Using pre-established static fields based on journal subject categories as proxies for such areas is generally a deficient strategy (Colliander, 2015). The categories are too broad, internally very heterogeneous (composed by different sub-fields that are substantially different), and they are to a large extent arbitrary. At the same time, journal subject categories are also intuitively in conflict with the dynamic grouping of research papers in temporal citation networks, networks that are self-organizing at the paper level and across arbitrary field delineations (e.g., van Raan, 2000; Scharnhorst et al., 2012).

The classification system we use is unique in relation to bibliometric analyses as it is based on direct citation relations between publications and not on the more arbitrary journal subject classifications. The use of a paper-based classification assumes that clustered papers are stronger related compared to a journal-based classification. If so the common problem in journal-based classifications of heterogeneity of the different fields, where subfields within a single field may differ significantly in terms of citation density, is to a large extent countered (Waltman & van Eck, 2012). The main features of this classification are the following:

- It is a paper-based classification. Publications are classified individually, thus avoiding the limitation of publications that are classified by the topic of the journals they are published in and not by their own content. Besides, the classification is based on articles and reviews only.
- It covers all eligible WoS-publications from 1993-2012, thus fully covering the whole period of this study (1993-2011).
- It is a hierarchical classification with three levels of disaggregation. There are 21 macro-fields that represent main scientific disciplines. These macro-fields contain themselves 784 different meso-fields, and finally we have a micro-classification composed by 21,167 micro-fields. All these levels have been used in our methodology for detecting breakthroughs in one way or another.
- Publications are assigned to one cluster only (at any level), thus avoiding the problem of multi-classification of publications and the subsequent multiplicative effects (Herranz & Ruiz-Castillo, 2012).

Notice, for consistency reasons (i.e. the fact that CoE from the humanities are excluded from this analysis) we have focused only on publications from the Science Citation Index and Social Sciences Citation Index (thus excluding the Arts & Humanities Citation Index). Also, for detection of breakthrough articles we exclude reviews as a document type and only consider research articles. Based on our previous definition of breakthroughs, review articles can hardly be defended as breakthroughs in a general sense as they mostly condensate and discuss the most recent and important developments and topics in a scientific domain, thus qualifying more as 'followers' than actual breakthroughs under our definition[5].

Consequently, and contrary to other related studies, we perform citation analyses on this large-scaled clustered network of journal articles in WoS.

*Advanced citation-based methods*

In our novel approach we operationalize three related conceptions of breakthrough research through three slightly different citation analysis methods. A very exclusive one where only the highest cited paper in a micro-cluster is considered as a candidate breakthrough paper (M1); as well as two conceptually different methods, one that detects candidate breakthrough papers among the two percent highest cited papers across all clusters according to the Characteristics Scores and Scales method (CSS) (Glänzel & Schubert, 1988) (M2a); and finally a more restrictive version of the CSS method where knowledge diffusion across macro clusters is also taken in as an extra parameter (M2b).

Characteristics Scores and Scales method

All benchmark identification approaches start with a first selection of the most cited publications within each of the 784 meso-fields. However, this selection is not based on the more traditional percentile approach where publications are ranked and top publications are selected from a percentile value within each field (cf. Waltman & Schreiber, 2013). For methods 2a and 2b, we use the so-called Characteristics Scores and Scales (CSS) method suggested by Schubert, Glänzel and Braun (1987) and demonstrated in other bibliometric approaches by Ruiz-Castillo (2012). The interesting possibilities of using CSS for research performance analyses are discussed by Glänzel, Thijs and Debackere (2013). The CSS method focuses on the common characteristics of citation distributions across fields and is based on the principle that citation distributions share some fundamental characteristics and similarities. The CSS method basically consists of the reduction of the original citation distribution to 'self-adjusting' classes by iteratively truncating the distribution to conditional mean values from the low end up to the high end.

In practice the method works as follows: taking the distribution of all publications in the WoS classified into the meso-categories, we calculate the mean of the number of citations of the distribution per meso-field ($\mu_1$); then we separate papers above and below the mean, and subsequently for the papers above the mean, we calculate a second mean (µ2), again we separate the papers above and below the µ2, and to those papers above the µ2 citation mean, we calculate a third mean (µ3), and finally we again separate these publications above and below µ3. As a result we can assign publications to four typologies which are described below. In addition to assigning publications to different typologies, as shown by Ruiz-Castillo (2012), this methodology also permits us to characterize the typologies according to the citations they receive. Table 1 presents the average values for the four typologies, as well as the proportion of citations they receive.

---

[5] However, they have been included in our CSS methodology, thus making the top classes of this methodology more exigent from a citation reception point of view.

**Table 1. Results of the CSS method applied to the 784 meso-fields**

|  | T1 (%) | T2 (%) | T3 (%) | T4 (%) |
|---|---|---|---|---|
| Proportion of publications in T1-T4 (sd in parenthesis) | 73.9 (2.6) | 18.9 (1.5) | 5.2 (0.8) | 2.0 (0.5) |
| Proportion of citations received by T1-T4 (sd in parenthesis) | 21.6 (1.8) | 32.1 (1.1) | 21.8 (0.9) | 24.5 (1.9) |

*T1-T4: Typologies 1 to 4.

The four typologies can be characterized as follows:

- Typology 1. *Lowly cited publications*: those that have an impact below the average of the entire field (µ1). They are the vast majority of publications in every field representing around 74 percent of all the publications and accounting for approximately 22 percent of all citations.

- Typology 2. *Moderately cited publications*: those that have an impact above the average of the entire field (µ1) but below the second mean (µ2). These publications represent approximately less than 19 percent of all the publications in their fields and receive 32 percent of all the citations in the field.

- Typology 3. *Highly cited publications*: these are publications that have an impact higher than µ2 but below µ3. They constitute approximately 5 percent of all publications within each meso-field and receive more than 21 percent of the citations in their respective fields.

- Typology 4. *Outstanding publications*. These publications represent barely 2 percent of all publications in every meso-field, but they alone receive around 25 percent of all citations in their meso-fields. In other words, these are the 2 percent most cited publications of every field and one in four citations given in their meso-fields goes to them.

There is a remarkable regularity across the fields of science and across the meso-fields in this study. This is very useful for our purposes as it allows us to apply the same approach across fields when we characterize the 'success' and 'dissemination' of the impact of these publications. A total of 16,250,505 publications (document types articles and reviews) covered by WoS during the period 1993-2012 and that have a meso-field in the CWTS classification[6] have been clustered in these four typologies. A total of 314,944 (1.9%) publications belong to type 4 (i.e. outstanding publications), of which 263,148 are of the document type 'article' (1.7% of all articles in the period). As a starting point, these 263,148 publications have been considered as potential breakthroughs of all sorts, yet a filtering approach is subsequently invoked. Notice, for all analyses we use a variable citation window.

Filtering out 'followers'

As discussed previously in our definition of breakthrough articles, being highly cited is in itself not sufficient to be considered a breakthrough, because publications should not be "*a mere follower of other highly cited publication(s)*" – it must have "*a genuine relevance on its own*". In this sense, it is not uncommon that highly cited publications are so simply because they have followed the steps of a previous breakthrough (or breakthroughs) and somehow profit from the 'spell' (i.e. innovativeness, novelty, relevance, etc.) of those previous publications. To operationalize a filtering out process of these 'followers' we have performed the following general steps:

- Identification of all pairs of potential breakthrough papers. Basically, we have identified potential breakthroughs citing another or other potential breakthrough(s). If we find such linkages, we label the citing breakthrough as B2 and the cited breakthrough as B1. Thus B2-papers are *potential* 'followers'[7].

- We then analyse the papers that cite B2 and check if they also cite B1, if so, we count these papers as double citers of B1 and B2.

- Finally, for B2 publications, we count how many of their citing papers that either simultaneously cites both B1 and B2 or only B2. Subsequently we enforce a threshold to designate 'followers'. Thus, for every potential 'follower' (i.e. B2 papers) *not* to be finally designated as a 'follower', the paper must have 70% or more of its citations alone (i.e. not being co-cited with B1 in more than 30% of its citations). While obviously arbitrarily defined, the main idea behind this threshold is that a breakthrough should not benefit too much from the spell of a previous potential breakthrough paper, as it should have a genuine impact on its own. Other values could be invoked, strengthening or diminishing this threshold.

---

[6] A total of 15,498,978 publications are classified as document type 'article' in the WoS in the same period.

[7] A limitation of this method is its reliance on the citation matching mechanism of the citation index used for the analysis. For example, breakthrough papers that have been published as pre-prints in repositories and that cannot be tracked back to their published version may cause problems in this step. Recently, the citation matching system of CWTS has been proved to be highly accurate (Olensky, Schmidt, & Van Eck, 2015), however this citation matching limitation must still be taken into consideration as it can signal some papers as breakthroughs while they are actually followers of other publications not properly matched as published cited documents.

We have applied this filter to the 263,148 'outstandingly' cited papers previously detected and 179,347 passed the filter of the followers being considered as potential breakthrough candidates (these papers are the basis for citation analysis methods 2a and 2b). The 179,347 breakthrough papers roughly correspond to 1% of the publications in WoS in the period 1993 to 2011.

What follows is a description of the three different citation analyses we have applied to detect potential breakthrough papers – including DNRF-breakthrough papers.

*Method 1: Breakthrough papers based on the micro-classification*

The first method (M1) is very simple but also extremely exclusive. It is based on the idea that the most cited paper of every micro-field can most likely be considered a breakthrough paper because it has the highest impact in it its micro-domain. This is a very restrictive definition of a breakthrough paper, because only *one* (or occasionally several) papers pass this filter. In fact, only 21,670 publications pass this filter as breakthroughs. For the time being, we ignore other potential dimensions of breakthrough and use this simple restrictive approach as a first general yardstick to compare with the other two methods.

*Method 2a: Breakthrough papers detected through the Characteristics Scores and Scales (CSS) method and filtering of 'followers'*

Methods 2a and 2b are related and conceptually different from M1. M2a is the most inclusive. This method is based on the 179,349 publications that have passed the filter of the 'followers'. We consider all these papers as potential breakthroughs. This method is the most inclusive (in terms of the number of publications finally considered as breakthroughs) of the three, and can work as an upper bound in the production of breakthroughs.

*Method 2b: Breakthrough papers detected through the Characteristics Scores and Scales (CSS) method, filtering of 'followers' and selecting those that have an impact in other different macro fields*

Based on the 179,349 publications considered for M2a, we have included a further parameter for determining breakthroughs. In this case, we introduce an interdisciplinary dimension in the delineation of breakthroughs assuming that substantial breakthroughs also have impact beyond their own micro- and meso-domains (i.e. they have an important outreach or knowledge diffusion to other major fields that in return cite these 'foreign' papers). This is the optimum approach in terms of identification of breakthroughs which meets all the criteria of our working definition. To operationalize this method, we have taken a relatively simple approach composed by the following steps:

- Taking all the citers of the 179,349 publications previously filtered, we counted the number of different macro-categories (i.e. a total of 21) from which they have received at least one citation.
- We then calculate the average of different external macro-fields where the breakthroughs of every meso-category have had some impact.
- Thus, based on the previous values, we consider a breakthrough paper as those publications within the same meso-category with an impact outside their own macro-field higher than the average of all the potential breakthroughs in the same meso-category.

Therefore, a breakthrough according to this third method (M2b) is a potential breakthrough paper that has an impact in more macro-categories than an average potential breakthrough within the same meso-category. A final set of 59,617 articles can be considered as breakthroughs according to this method, approximately 0.38% of all the articles in the time period analysed.

*Multi-layered case study evaluation of the three citation-based approaches*

Obviously, a technical procedure of identifying and filtering out highly cited papers is not sufficient to claim that such papers are indeed breakthrough papers or that they are a proxy for some breakthrough research, even though some theory may suggest that extremely highly cited papers could be a sign of such breakthrough research (e.g., www.highlycited.com; van Noorden, Maher & Nuzzo, 2014). It is important to recognize that while we may postulate this in general, individual differences are to be expected. Also, it is important to acknowledge that breakthrough research nowadays is most likely not disseminated in one paper. This could be the case, for example, for some methodical or instrumental breakthroughs, but in many other instances breakthrough research is developed and disseminated in a succession of papers and far from all of them will become highly cited. It is therefore more appropriate to examine the approaches in relation to an aggregate unit of analysis, i.e., a unit with a certain publication portfolio. So in order to examine the usefulness of our proposed methodology we need to examine the results and compare them to some qualitative yardstick. We need to know whether some units of analysis indeed have produced breakthrough research as judged by peers and to what extent such units can be identified through our proposed citation-based methods.

The ability of the methods to identify potential breakthrough research is explored in three case studies. In the most extensive analysis we examine publications coming from the CoE funded by the DNRF. To substantiate this analysis, we also examine the performance of the proposed citation approaches in relation to a set of publications linked to markedly different funding instruments, i.e. small grant types from the DFF. Finally, and not related to the first two case studies, we also make a simple performance analysis where we compare the approaches to a recent algorithmic attempt at predicting breakthrough papers (Ponomarev et al., 2014b).

Since 1993 the DNRF has funded CoE in Denmark. The foundation was explicitly set up to identify and fund potentially excellent or breakthrough research. Potential excellence is demanded from applicants. The selection process is hard with high rejection rates, but the eventual funding of the centres is substantial and long-termed, up to 10 years[8]. From a science policy perspective, the rationale is that these conditions represent a unique setting for the CoE to do outstanding research. However, selection and funding of potentially excellent research is risky as not all funded CoE will perform equally well, or produce breakthrough research for that matter.

For the present evaluative purposes, analysing CoE is interesting exactly because their main purpose is to produce excellent or breakthrough research. Obviously, normative standards for when such an objective is reached is impossible to set up, but we can expect individual CoEs to fulfil this objective to varying degrees. The centres are created and formed around a specific foundational research problem. We can therefore expect that numerous papers from a CoE more or less address the same overall research problems. If we then examine the performance of the CoE as a unit, we may be able to say something about the potential breakthroughs in the research problems they set out to investigate.

---

[8] Funding rounds happen with 3 to 5 year intervals, the selection process has two steps and around 6% of the intial proposals are funded; many CoEs receive 6 to 9 million Euros, but some receive up to 15 million Euros.

The validated publication data and related information we have on the CoE comes from an international panel evaluation of the DNRF as a funding instrument (Schneider & Costas, 2013)[9]. In relation to that, the DNRF was asked beforehand by the commissioners of the evaluation whether they could designate some CoEs who were known to have produced breakthrough research of some kind. It is important to emphasize that the result of this exercise was not revealed to us until after the three citation-based methods had been developed. We received the results afterwards in order to evaluate the proposed methods. As it turned out, the DNRF provided a list of eight CoE whom they considered to have produced breakthrough research in the period analysed. Notice, the list is not exhaustive; some of the 66 CoE were still running in 2011, the final publication year included in the study, and others could probably have been chosen (the assessment assignment did not stipulate exhaustiveness). However, for the present purposes this is irrelevant. Having eight CoE is sufficient in order to examine the usefulness of the proposed methodology. For the validation, each of the eight chosen CoE provided a small description of their scientific contribution and indicated a number of papers they themselves considered important.

We find this information extremely important for the validation purposes, but, we cannot rule out that the selection of CoE and/or the designation of important papers are influenced by bibliometric indicators in the first place, and if so, selection bias is present. Further, among the designated CoE, some of the indicated important papers were published outside the period analysed in this study. One CoE actually indicated important papers that were not included in the original data set because the specific publication year in question was missing in the original publication list. Lastly, one of the eight chosen CoEs was still active at the time and for the present analysis has only been active in 4-5 years. The latter fact significantly lowers the probability of detecting breakthrough papers for this particular CoE.

While in principle all CoE, or other units of analysis, can produce what turns out to be a potential breakthrough research papers according to one or several of the suggested approaches, we use frequency of such articles within a unit (CoE) as a criterion and an indication of the robustness of the findings. Depending on the approach, a unit with more candidate papers is more likely to have produced actual breakthrough research.

---

[9] http://fivu.dk/en/publications/2013/evaluation-of-the-danish-national-research-foundation.

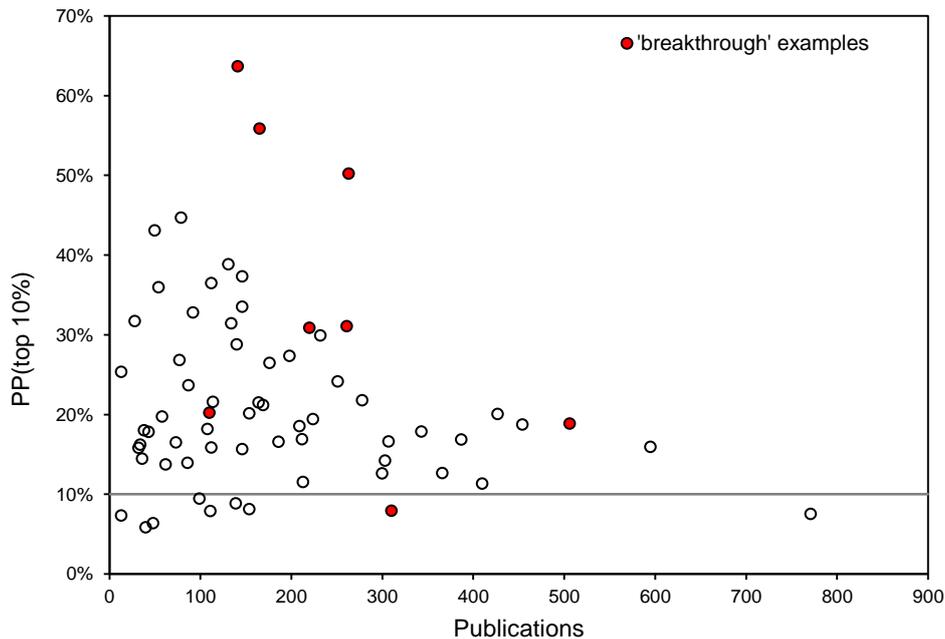

**Figure 1: PP(top 10%) for all 66 CoEs, dark dots are the 8 CoE chosen as examples of 'breakthrough' research; these CoE are used to assess the applied methodology.**

Figure 1 above shows the overall PP(top 10%)[10] indicator for all 66 CoE included in the analysis as a function of publication output for the whole period 1993 – 2011. The dark dots in Figure 1 mark the eight CoE selected as examples of centres that have produced breakthrough research as judged by the peers. Noticeable, the three highest performing CoE are among the chosen examples, but also CoE from other strata of the impact distribution. Since our approach focuses upon detecting potential breakthrough papers among the set of extremely highly cited publications, CoE with several such publications will most likely also be among the highest performing CoE, whereas the challenge will be further down the impact distribution.

Given their rationale, CoE are a very appropriate unit of analysis for the present purposes; however, to substantiate the validation process, we also compare the combined results for the aggregate set of DNRF publications from 2005 to 2011 (i.e. merging the CoE publications for this period) to that of another Danish funding institution, the DFF. A recent evaluation of publications linked to grants from DFF in the period 2005 to 2008 (Schneider et al., 2014)[11] enables a specific comparison between the two funding instruments when it comes to producing potential breakthrough articles. While both publication sets are of approximately similar size when it comes to publications from 2005 to 2011, the two units have quite different modes and funding instruments of widely different scales. DFF grants are generally smaller, more individually oriented towards younger researchers, and have a considerably shorter duration than centres funded by the DNRF[12]. Essentially, these instruments play different roles in the Danish research funding ecology. Consequently, due to specific attention to excellence, we would expect that a relatively larger proportion of articles in the DNRF set end up as potential breakthroughs compared to the set of publications funded by the DFF. As a final control, we also compare the two aggregate funding sets, DNRF and DFF, to the total set of publications from Denmark in the given period.

---

[10] PP(top 10%) refers to the proportion of publications of a unit within the 10% most cited papers of their fields.

Finally, the proposed citation based approaches to identifying potential breakthrough articles are also explored in relation to the findings in Ponomarev et al. (2014b) where two models for early detection of 'candidate breakthroughs' also based on citation analyses are proposed. Consequently, we evaluate the three citation approaches and their ability to detect potential breakthrough papers in a multi-layered case study where we:

1. Examine the number of potential breakthrough papers for each of the 66 individual CoEs. Eight of these CoE have been assessed as having produced breakthrough research, hence we specifically examine to what extent these eight CoE have produced potential breakthrough papers. We would expect some or all of these CoE to have an ample number of breakthrough papers compared to the remaining CoE.

2. Compare the combined set of all publications coming from the 66 CoE in the period 2005 to 2011 to a corresponding set of all publications linked to various grant types from the DFF, to substantiate the first analysis. As these grants are substantially different from the CoE instrument, we would expect to see fewer potential breakthrough papers coming from such grants and thus the DFF set of publications.

---

[11] http://ufm.dk/publikationer/2014/filer-2014/analyses-of-the-scholarly-and-scientific-output-from-grants-funded-by-the-danish-council-for-independent-research-from-2005-to-2008.pdf.

[12] Success rates are around 10% for DFF applications and grant typically run for two to four years with a few exceptions.

3. Finally, we examine to what extent the three citation-based approaches are able to detect the supposedly 11 known breakthrough publications in molecular biology and genetics examined in Ponomarev and colleagues (2014b).

The next section presents the results of the methods and the subsequent validations.

## Results

In this section we present the main results of the three different methods. As stated above, we focus upon publications linked to the individual CoE in the period 1993 to 2011, as well as aggregated set CoE publications linked DNRF in the period 2005 to 2011[13]. Based on the delineations of breakthroughs previously specified in methods 1, 2a and 2b, we study the presence of these types of publications in the various publication sets. We present the identified breakthrough papers in three science maps. It is the same base map showing the 784 meso-fields which is used in the three analyses. The meso-fields include the micro-fields and they are positioned in relation to each other according to citation links. In these maps, we indicate potential breakthrough papers from DNRF as numbers placed on top of the meso-field (i.e. numbers are the actual number of DNRF breakthrough papers in the micro-field).

Figure 2 below shows the frequency distributions of breakthrough papers for the 66 CoE for each of three methods.

---

[13] For the CoE analysis, 10,803 publications were eligible; for the aggregate DNRF analysis with a shorter publication window, 7108 publications were eligible.

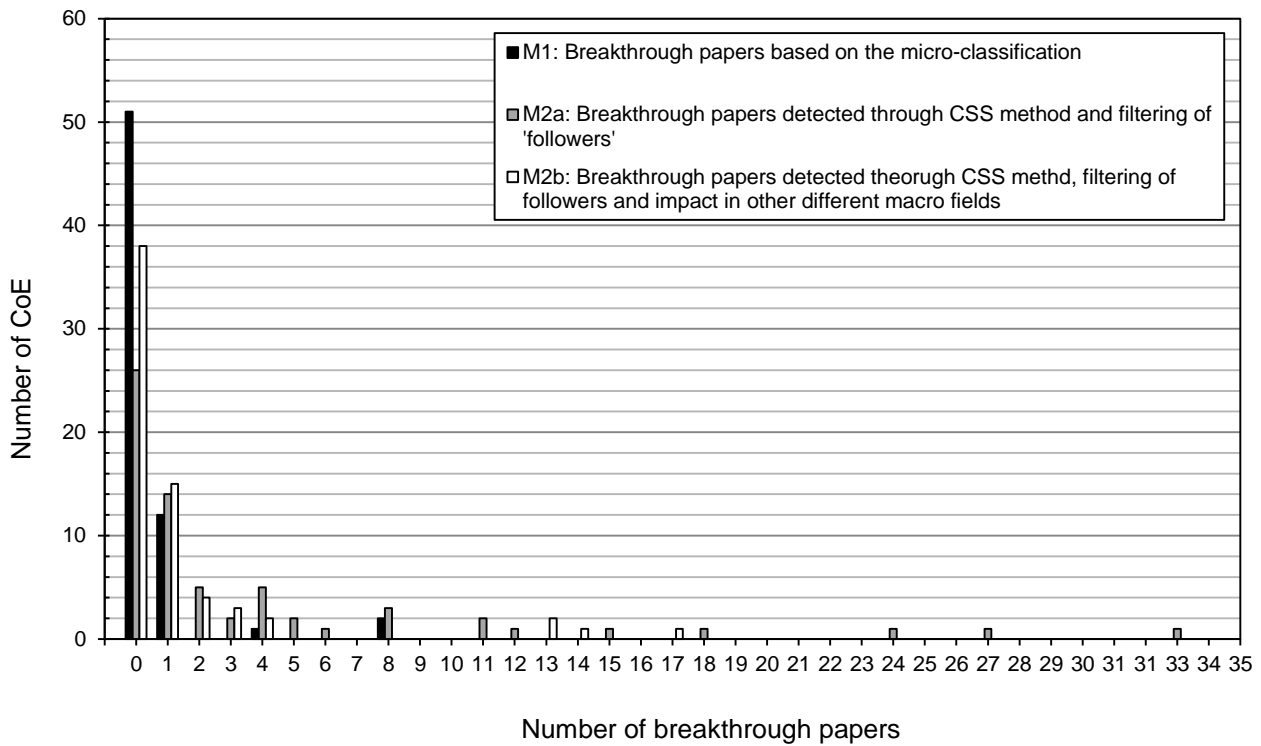

**Figure 2: Number of detected candidate breakthrough papers according to Centres of Excellence for the three different citation-based methods.**

*Results of Method 1: 'Breakthrough' papers based on the micro-classification*

Method 1 is the most simple but also the most restrictive of the three methods. A breakthrough paper is here defined as the paper with the highest impact in the micro-field. Such a paper is considered to have tremendous importance or constitution for the research area. In all, 21,167 micro-fields have been established and 21,670 publications pass the filter as breakthroughs (i.e. the number is higher than the number of micro-fields because there are ties among the highest cited papers in some fields). Of these 21,670 potential breakthrough papers, 0.15% or 32 papers come from the set of DNRF-publications. Notice, the set of DNRF-publications constitutes only 0.07% of all publications analysed, thus there is an overrepresentation of potential breakthrough papers from the set of DNRF-publications, i.e. a ratio of 2.4. Also, the 32 potential breakthrough papers constitute 0.3% of the papers in the DNRF-publication set.

Table A1 in the appendix provides the distribution on potential breakthrough papers from Method 1 identically to the result depicted in the first histogram in Figure 2. It is clear that the distribution shows the characteristic skewness found in most scientometric distributions: three CoEs account for 20 of the potential breakthrough papers. Whereas 12 CoE have one breakthrough paper and 51 CoE have no breakthrough papers according to this definition.

Figure 3 below shows the science map where the potential breakthrough papers are marked with a number on top of the meso-fields (circles). Notice, the breakthroughs are detected in the micro-fields, and a meso-field is constituted by several micro-fields. We have identified the major research fields with colors and the size of circles indicates the relative size of the meso-fields.

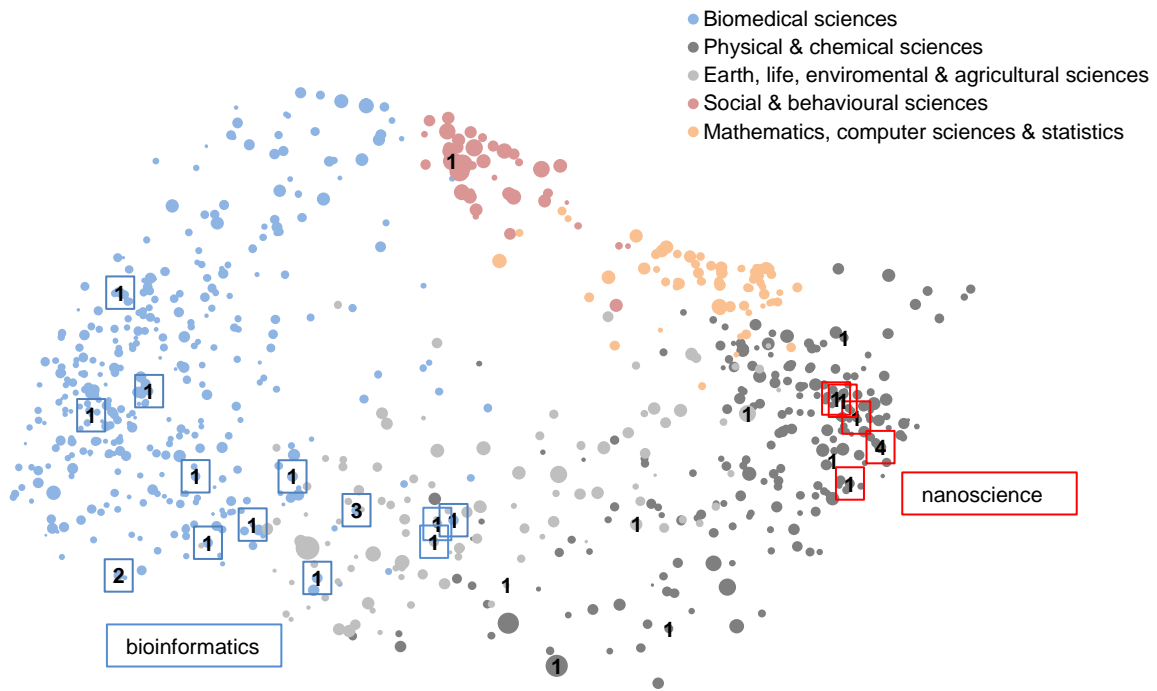

**Figure 3: Method 1 – map of 784 meso-fields where potential DNRF-'breakthrough' papers are indicated with numbers on top of fields.**

As stated above, three CoE account for 20 of the potential breakthrough papers. As indicated on the map, 12 of them concern breakthrough research in bioinformatics and eight of the papers, what we today would call nanoscience. From Table A1 we can see that the three CoE accounting for two-thirds of the breakthrough papers are all identified by the peer review as centres that had produced breakthrough research. In total, five out of the eight identified CoE have at least one candidate breakthrough paper in method 1, three of them ranked in top 3.

*Results of Method 2a: 'Breakthrough' papers detected through the 'Characteristics Scores and Scales' (CSS) method and filtering of 'followers'*

Method 2a is based on the CSS method where so-called 'followers' are filtered out. Even though, the method only considers the 2% most cited papers, it is the least restrictive method of the three we explore in this study. In total, 179,349 publications from 1993-2011 have been detected as potential breakthroughs, of these 241 come from the set of DNRF-publications. This corresponds to 0.13% of all potential 'breakthrough' papers defined by method 2a, i.e., a slightly lower ratio of 1.9 compared to M1, and 2.2% of the papers in the set of DNRF-publications.

Figure 2 and Table A2 in the appendix show the distribution of breakthrough papers among the 66 CoE. Again the distribution is skewed, for slightly more than one third of the CoE (26), no potential breakthrough papers are detected. Nevertheless, as expected many CoE (40) now have at least one potential breakthrough paper according to this method. It is also expected that the three CoE which together accounted for two-thirds of the potential breakthrough papers in M1 are also highly visible in this method given its lesser restrictions. Together they account for one-third of the potential breakthrough papers in Method 2a. However, there are new candidates emerging with this method. The second highest number of potential breakthrough publications, 27, is from a CoE that had one potential breakthrough publication in Method 1. In all, six out of the eight assessed CoE have at least one candidate breakthrough paper in Method 2a, five of them ranked among the top 8.

Figure 4 below shows the distribution of potential breakthrough papers identified by Method 2a over the 784 meso-fields. In this map we have marked the 'newcomer' which concerns register-based epidemiological research. Bioinformatics and nanoscience papers can be located in the same areas or close by as in the previous map in Figure 3. Breakthrough papers in areas such as catalysis, metal structures, as well as muscle and sensory-motor research, are only indicated with their numbers for overview reasons.

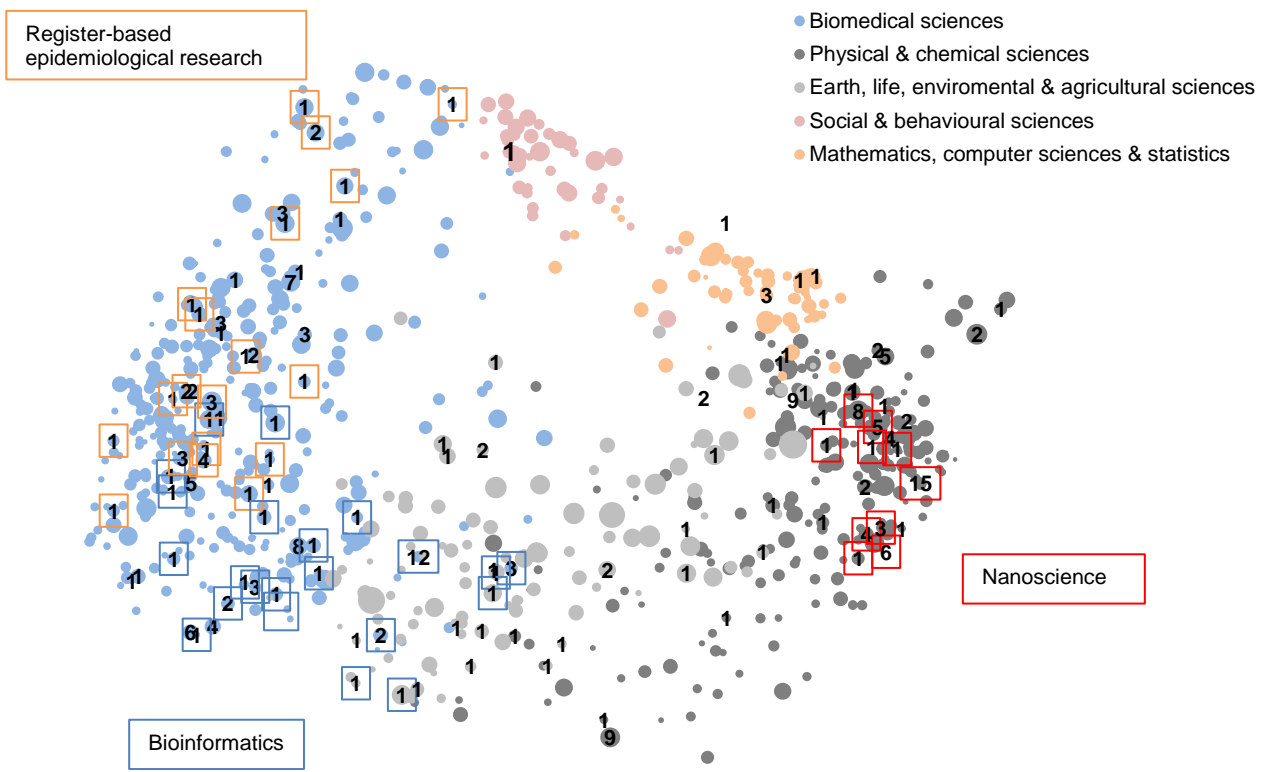

**Figure 4: Method 2a – map of 784 meso-fields where potential DNRF-'breakthrough' papers are indicated with numbers on top of fields.**

*Results of Method 2b: 'Breakthrough' papers detected through the 'Characteristics Scores and Scales' (CSS) method, filtering of 'followers' and selecting those that have an impact in other different macro fields*

The third and final method, M2b, is much more restrictive than M2a because it further delimits the set of potential 'breakthroughs' into only those publications that have an impact in more macro-categories than an average potential 'breakthrough' paper within the same meso-category. Hence, we try to reflect the potential knowledge diffusion of the potential 'breakthroughs'. A total of 59,617 articles are considered potential 'breakthroughs' according to this method. Of these, 0.16% or 97 papers come from the set of DNRF-publications and this corresponds to approximately 1% of the DNRF-publications, i.e., a ratio of 2.3 considering that DNRF-publications constitute 0.07% of the total investigated.

Figure 2 and Table A3 in the appendix show the distribution of breakthrough papers among the 66 CoE. Again we see a skewed distribution. Not surprisingly given the requirements of interdisciplinary citations from other macro-categories, the number of potential breakthrough papers is markedly reduced compared to Method 2a and the number of CoE where no potential breakthrough papers could be identified has increased considerably to 57% of all CoE included. It is noteworthy that the number of potential breakthrough papers is significantly higher for the top four CoE compared to the other CoE. Centres ranked 5-6 have four potential breakthrough papers, whereas the CoE ranked 3-4 have 13 breakthrough papers each. In total, five out of the eight assessed CoE have at least one candidate breakthrough paper in Method 2b, five of them among the top 8.

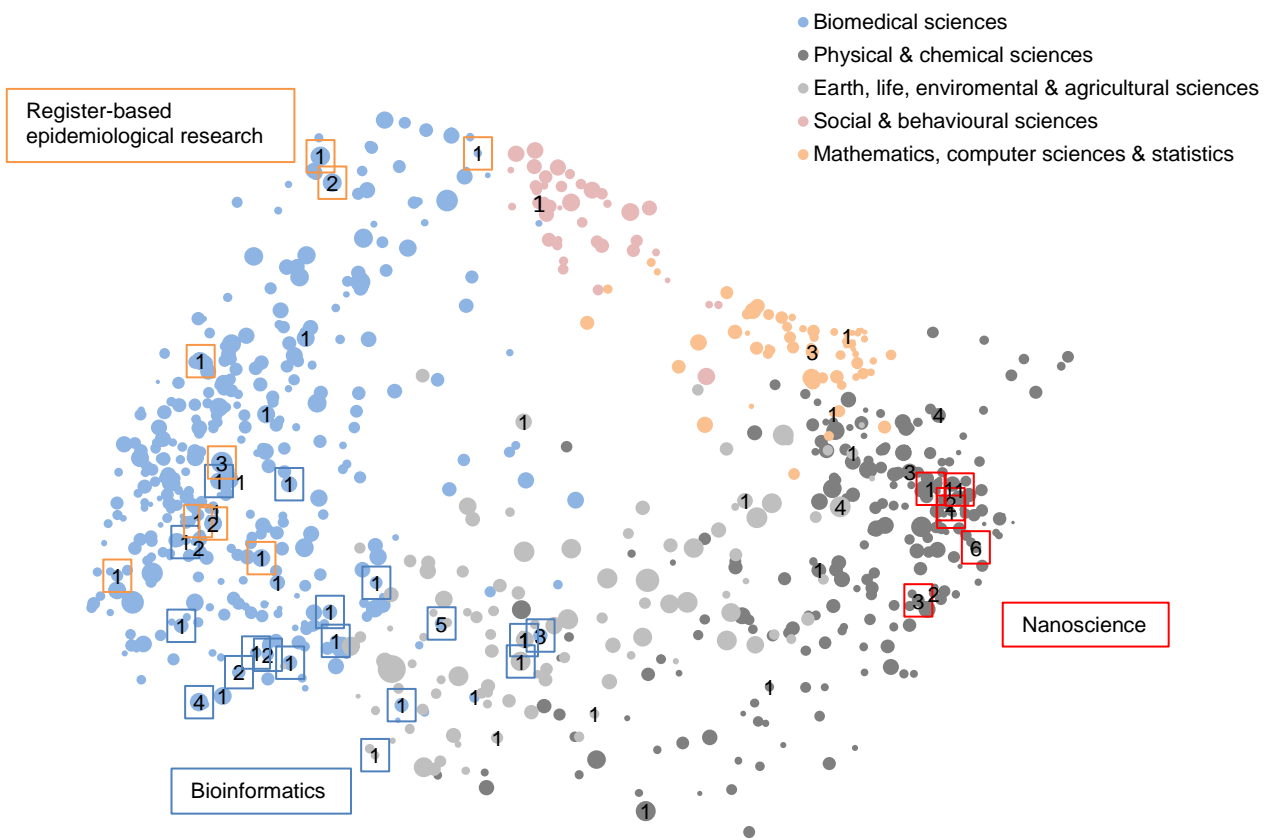

**Figure 5: Method 2b – map of 784 meso-fields where potential DNRF-'breakthrough' papers are indicated with numbers on top of fields.**

Figure 5 above shows the distribution of potential 'breakthrough' papers identified by Method 2b over the 784 meso-fields. The four CoE together account for 59% of the potential 'breakthrough' papers identified with M2b. Not surprisingly, it is the same three-four CoEs and research areas as depicted in the other two maps which are most visible: bioinformatics (two CoEs), nanoscience and register-based epidemiological research.

*Comparison between publications linked to Centres of Excellence funded by the DNRF and publications linked to grants from the Danish Council for Independent Research in the period 2005 to 2011.*

As a result of a recent evaluation of the Danish Council for Independent Research (DFF), we have a validated publication set of WoS journal articles published between 2005 and 2011 which is linked to DFF-grants awarded in the period 2005 to 2008 (Schneider et al., 2014). This publication set is a subset of the overall data set of publications from 1993 to 2011 used in the present analyses. Similarly, we are able to restrict the CoE publications to those published between 2005 and 2011 and merge them into a combined aggregate set of DNRF publications. In this way, we are able to compare the incidence of potential breakthrough papers detected with the three citation-based methods between the two aggregated publication sets. Due to the different aims and modes of the two funding instruments, we generally expect more frequent incidences of detected breakthrough papers in the DNRF set due to the specific attention given to 'excellence' by this funding instrument. We also include a third publication set comprising all Danish publications in the period 2005 to 2011 as a common reference point for the two other sets. We examine the results as totals for the two funding bodies and the Denmark. The results are presented in Table 2 below.

**Table 2. Proportion of Danish, DFF and DNRF 'breakthrough' articles from 2005 to 2011 (p = publications).**

|  | 'Breakthrough' methods | | |
| --- | --- | --- | --- |
|  | Method 1 | Method 2a | Method 2b |
| p 'breakthrough': Database | 1,369 | 13,997 | 4,326 |
| p 'breakthrough': Denmark | 27 | 292 | 106 |
| p 'breakthrough': DFF | 1 | 24 | 10 |
| p 'breakthrough': DNRF | 5 | 48 | 20 |
| % p 'breakthrough' in own set: | 0.03% | 0.38% | 0.14% |

|  | 'Breakthrough' methods | | |
|---|---|---|---|
|  | Method 1 | Method 2a | Method 2b |
| Denmark (n = 77571) % p 'breakthrough' in own set: DFF (n = 6205) | 0.02% | 0.39% | 0.16% |
| % p 'breakthrough' in own set: DNRF (n = 7108) | 0.07% | 0.68% | 0.28% |
| % DFF p among Danish 'breakthroughs' | 3.7% | 8.2% | 9.4% |
| % DNRF p among Danish 'breakthroughs' | 18.5% | 16.4% | 18.9% |
| % Danish p among database 'breakthroughs' | 1.97% | 2.09% | 2.45% |

The first four rows of Table 2 present the actual number of identified potential breakthrough papers in the three categories for the database in the restricted period from 2005 to 2011 examined, for the Danish set of publications, for the DFF set of publications and for the DNRF set of publications. Notice, there are duplicates between the categories. Papers identified by a more restrictive method will most probably turn up in one or both of the lesser restrictive methods (2a, 2b); but there are exceptions. Also notice that the publication window is from 2005 to 2011 and we apply the usual variable citation window. The subsequent rows present the proportion of potential breakthrough papers identified.

If we first compare the profiles across the three methods for the proportion of potential breakthrough papers relative to the size of the individual publication sets in each method, it is evident that the Danish and DFF sets of publications are very similar in profile, whereas the DNRF-set is very different with roughly twice as many candidate breakthrough papers, relatively speaking. What should be noticed is that the highest discrepancy is in the first and most restrictive method; here the DNRF set clearly outperforms the sets from the DFF and Denmark. If we then examine the relative share of DNRF and DFF publications among the breakthrough papers for Denmark, we again see that the DNRF set produces twice as many candidate papers in M2a and M2b, and noticeably more than four times as many in M1, the most restrictive method.

The results confirm what we expected. Markedly more potential breakthrough papers are linked to the aggregate DNRF funding instrument. The DNRF fund CoE with the explicit goal of producing 'excellent' or 'breakthrough' research. The DNRF tune their funding instrument for that purpose, so if very highly cited articles detected through contextualization and filtering can be seen as proxies for such research efforts, we should expect such a unit of analysis to outperform the control groups. It is also notable that the most marked difference between the DNRF publication set and the controls is in the most restricted method (M1) where only the highest cited defining article of a micro-field is selected. To a certain extent this corroborates the notion that breakthrough research is actually produced and even defines or establishes specific and new fields, and it also suggests that the funding instrument to some extent serves its task. Here we should remind that the willingness to make risk in funding is important as far from all CoEs end up producing sustained breakthrough research.

As a final validity check, we also compared our findings to the 11 known breakthrough publications in molecular biology and genetics examined in Ponomarev and colleagues (2014b). Seven of the papers were detected by at least one of our methods. Two papers were detected by all three methods and two others were detected by M2a and M2b, whereas three papers were detected alone by M2a, the least restrictive. Obviously, our definitions and operationalisation matter here and they can be discussed. In fact, the four papers which our methods do not detect are not among the two percent most cited and this of course raises the question to what extent papers disseminating potential breakthrough research end up as highly cited, clearly not all do.

## Discussion and reflections

We have presented three advanced citation-based methods aimed at detecting potential breakthrough research papers. We have discussed the premises and assumptions behind the methods and examined to what extent very highly cited articles when contextualized through direct-link clustering and filtered for potential 'followers' can be seen as proxies for breakthrough research. Overall we find the advanced citation methods promising. In our validation studies we generally find that contextualizing and filtering of very highly cited papers do point to breakthrough or important research findings. When we compare our findings to the eight examples provided by the DNRF, we can conclude that potential breakthrough papers are identified for four CoE in all three methods; potential breakthrough papers are identified for five CoE in two or more of the methods, and potential breakthrough papers are identified for six CoE in one or more of the methods. Notice, the three clearly highest ranked CoE according to PP(top 10%), shown in Figure 1, are also the three most prominent in the breakthrough analyses, multiple articles from these CoE are detected by all three methods. While it is not entirely surprising that the major findings, i.e. the same three to four research areas and CoE, recur in the results of the three methods, after all the methods are variations over the same idea, it is noticeable that these areas come out so strong.

Three CoE have several highest cited publications across the micro-fields. They all have numerous potential breakthroughs among the highest cited publications and a considerable number of their potential breakthrough papers receive citations from a more than average number of other macro-categories. The latter indicating the widespread diffusion and use of this particular research. Consequently, and not surprisingly, our methods correlate well with very high performance of the units of analyses. Obviously, having an amassed unit of analysis such as a centre makes the methods much more robust compared to merely identifying isolate candidate breakthrough papers. We find it reasonable to assume that the likelihood for identifying real breakthrough research is considerably higher with groups producing multiple candidate papers.

Also noticeable is the fact that for two examples we did not identify any potential breakthrough papers. Of the two examples, one was still active and its publication activity for the period under investigation is limited, hence not identifying potential breakthrough papers in this case can be linked to an effect of time. This explanation does not hold for the other case. Here the funding of the CoE expired within the period under investigation.

When we compared our findings to the 11 assumed breakthrough papers in Ponomarev et al (2014c) we likewise miss out on four articles. A common explanation for these misses are the citation and contextual statuses of the these papers in our approach; both those belonging to the CoE, as well as the four papers in Ponomarev et al (2014c) are not considered very highly cited given our definition and thresholds. Indeed, looking at the citations received by these four articles, then it is apparent that they are not very highly cited given the context of the publication (i.e. journals such as *Nature*, *Science* and *Gene*) compared to the other seven in the sample actually detected. In as much as these articles do present breakthrough research, and the same goes for the CoE where the overall citation performance was not outstanding, then it is obvious that citation-based methods alone have clear limitations. Given the parameters and thresholds, such methods only focus on the most significant signals in a citation distribution, i.e. the top percentiles. As signals become weaker going down the distribution, the methods are not able to detect such papers and given our definition and operationalization, therefore not able to identify potential breakthrough papers.

Of an interesting note, it is well known that the aggregate Journal Impact Factor is a vague predictor for individual article citation scores (e.g., Seglen, 1992). It is also well-known that broad high-impact journals such as *Science*, *Nature*, *Cell*, *Physical Review Letters, New England Journal of Medicine, British Medical Journal* and the like, are very prestigious to publish in although difficult to get into. It is also recognized that these journals are selective and self-serving in their topics. Nevertheless, these are the predominant journals producing candidate breakthrough papers in our analyses.

To conclude, we have tried empirically to detect potential breakthrough papers and used them as proxies for breakthrough research. If citation signals turn out to be strong, the methods are, other things being equal, able to detect them and this we have done in this analysis. As we have argued the methods have numerous limitations. Particularly we would like to emphasize that we are only able to detect strong signals through citation analysis. While something will go undetected, we do find that the advanced citation-based methods, especially at the aggregate level of analysis (e.g. centres or larger publication sets) are a very useful tool for sorting and characterizing among highly cited papers and high performing units. The methodology is simple and replicable and there is consistency among the methods and the results, especially among the strongest signals. But the results are also consistent to the examples suggested by DNRF themselves. Our example where different funding instruments lead to different prediction when it comes to number of breakthrough papers illustrates the usefulness of the approach as a rudimentary measure of 'breakthrough research'. Indeed, one could argue that the methods are too restrictive, but identifying important papers which are located lower down in the citation distribution is very difficult as signals get nosier. Finally, we should state that there may be a time-lag effect in our approach. Citation analyses are per definition retrospective. Of the 26 CoE where we did not detect any potential breakthrough papers, 19 were still active during the period under investigation. Hence, it is reasonable to suggest that some perhaps most of them will eventually produce potential breakthroughs it just takes some time for them to be visible. In that respect we claim that the presented methods to identify potential breakthrough research by use of refined citation analyses is an important extension to traditional citation and impact analyses.

# Appendix

**Table A1. Distribution of breakthroughs by Centers of Excellence and the total DNRF – Method 1.**

| CoE no | No. of publications from CoE | No. of Breakthrough papers (M1) | Percentage of breakthrough papers from CoE (%) |
|---|---|---|---|
| 12* | 201 | 8 | 3.9 |
| 6* | 256 | 8 | 3.13 |
| 38* | 121 | 4 | 3.31 |
| 11 | 387 | 1 | 0.26 |
| 2 | 578 | 1 | 0.17 |
| 31 | 95 | 1 | 1.05 |
| 32* | 148 | 1 | 0.68 |
| 17 | 139 | 1 | 0.72 |
| 67 | 100 | 1 | 1.0 |
| 14* | 484 | 1 | 0.21 |
| 19 | 13 | 1 | 7.69 |
| 10 | 239 | 1 | 0.42 |
| 34 | 57 | 1 | 1.75 |
| 4 | 310 | 1 | 0.32 |
| 48 | 205 | 1 | 0.49 |
| DNRF Total | 10,804 | 32 | 0.30 |

* In the sample of eight CoE judged to have produced breakthrough research based on peer evaluation.

**Table A2. Distribution of breakthroughs by Centers of Excellence and the total DNRF – Method 2a.**

| CoE no | No. of publications from CoE | No. of Breakthrough papers (M2a) | Percentage of breakthrough papers from CoE (%) |
|---|---|---|---|
| 6* | 256 | 33 | 12.9 |
| 2 | 578 | 27 | 4.7 |
| 38* | 121 | 24 | 19.8 |
| 12* | 201 | 18 | 9.0 |
| 14* | 484 | 15 | 3.1 |
| 40 | 241 | 12 | 5.0 |
| 15 | 736 | 11 | 1.5 |
| 32* | 148 | 11 | 7.4 |
| 35 | 114 | 8 | 7.0 |
| 10 | 239 | 8 | 3.3 |
| 11 | 387 | 8 | 2.1 |
| 4 | 310 | 6 | 1.9 |
| 37 | 266 | 5 | 1.9 |
| 56 | 78 | 5 | 6.4 |
| 34 | 57 | 4 | 7.0 |
| 5 | 281 | 4 | 1.4 |
| 54 | 163 | 4 | 2.5 |
| 16 | 104 | 4 | 3.8 |

| CoE no | No. of publications | No. of Breakthrough papers | Percentage |
|---|---|---|---|
| 36 | 182 | 4 | 2.2 |
| 44* | 245 | 3 | 1.2 |
| 83 | 121 | 3 | 2.5 |
| 67 | 100 | 2 | 2.0 |
| 68 | 109 | 2 | 1.8 |
| 80 | 48 | 2 | 4.2 |
| 48 | 205 | 2 | 1.0 |
| 17 | 139 | 2 | 1.4 |
| 13 | 81 | 1 | 1.2 |
| 33 | 34 | 1 | 2.9 |
| 19 | 13 | 1 | 7.7 |
| 31 | 95 | 1 | 1.1 |
| 43 | 405 | 1 | 0.2 |
| 46 | 72 | 1 | 1.4 |
| 55 | 161 | 1 | 0.6 |
| 51 | 86 | 1 | 1.2 |
| 8 | 25 | 1 | 4.0 |
| 82 | 111 | 1 | 0.9 |
| 58 | 141 | 1 | 0.7 |
| 62 | 393 | 1 | 0.3 |
| 94 | 50 | 1 | 2.0 |
| 97 | 121 | 1 | 0.8 |
| DNRF Total | 10,803 | 241 | 2.23 |

*In the sample of eight CoE judged to have produced breakthrough research based on peer evaluation.

**Table A3. Distribution of breakthroughs by Centers of Excellence and the total DNRF – Method 2b.**

| CoE no | No. of publications from CoE | No. of Breakthrough papers (M2b) | Percentage of breakthrough papers from CoE (%) |
|---|---|---|---|
| *38 | 121 | 17 | 14.0 |
| *12 | 201 | 14 | 7.0 |
| 2 | 578 | 13 | 2.2 |
| *6 | 256 | 13 | 5.1 |
| 35 | 114 | 4 | 3.5 |
| 4 | 310 | 4 | 1.3 |
| 40 | 241 | 3 | 1.2 |
| 14* | 484 | 3 | 0.6 |
| 54 | 163 | 3 | 1.8 |
| 80 | 48 | 2 | 4.2 |
| 15 | 736 | 2 | 0.3 |
| 36 | 182 | 2 | 1.1 |
| 34 | 57 | 2 | 3.5 |
| 19 | 13 | 1 | 7.7 |
| 37 | 266 | 1 | 0.4 |
| 31 | 95 | 1 | 1.1 |
| 32* | 148 | 1 | 0.7 |

Accepted for publication in Journal of the Association for Information Science and Technology

| | | | |
|---|---|---|---|
| 33 | 34 | 1 | 2.9 |
| 17 | 139 | 1 | 0.7 |
| 10 | 239 | 1 | 0.4 |
| 43 | 405 | 1 | 0.2 |
| 46 | 72 | 1 | 1.4 |
| 48 | 205 | 1 | 0.5 |
| 56 | 78 | 1 | 1.3 |
| 94 | 50 | 1 | 2.0 |
| 67 | 100 | 1 | 1.0 |
| 68 | 109 | 1 | 0.9 |
| DNRF Total | 10,803 | 97 | 0.9 |

*In the sample of eight CoE judged to have breakthrough research based on peer evaluation.